\documentclass[referee,sn-basic]{sn-jnl}
\usepackage{geometry}
\usepackage{graphicx}%
\usepackage{caption} 
\usepackage{subcaption}  
\usepackage{multirow}%
\usepackage{amsmath,amssymb,amsfonts}%
\usepackage{amsthm}%
\usepackage{mathrsfs}%
\usepackage[title]{appendix}%
\usepackage{xcolor}%
\usepackage{textcomp}%
\usepackage{manyfoot}%
\usepackage{booktabs}%
\usepackage{algorithm}%
\usepackage{algorithmicx}%
\usepackage{algpseudocode}%
\usepackage{listings}%

\newcommand{\RR}{\mathbb{R}}
\newcommand{\LL}{\mathcal{L}}

\newcommand{\vecf}{\vec{f}}

\newcommand{\vecu}{\vec{u}}
\newcommand{\vecv}{\vec{v}}

\newcommand{\ep}{\varepsilon}
\newcommand{\mc}{\mathcal}

\renewcommand{\vec}[1]{\mathbf{#1}}

\usepackage{xcolor}

\begin{document}

\title[Representing stimulus motion with waves in adaptive neural fields]{Representing stimulus motion with waves in adaptive neural fields}

\author[1]{\fnm{Sage} \sur{Shaw}}\email{sage.shaw@colorado.edu}

\author[1,2]{\fnm{Zachary P} \sur{Kilpatrick}}\email{zpkilpat@colorado.edu}

\affil[1]{\orgdiv{Department of Applied Mathematics}, \orgname{University of Colorado Boulder}, \orgaddress{ \city{Boulder}, \state{CO}, \country{USA}}}

\affil[2]{\orgdiv{Institute for Cognitive Sciences}, \orgname{University of Colorado Boulder}, \orgaddress{\city{Boulder}, \state{CO}, \country{USA}}}

\abstract{Traveling waves of neural activity emerge in cortical networks both spontaneously and in response to stimuli. The spatiotemporal structure of waves can indicate the information they encode and the physiological processes that sustain them. Here, we investigate the stimulus-response relationships of traveling waves emerging in adaptive neural fields as a model of visual motion processing. Neural field equations model the activity of cortical tissue as a continuum excitable medium, and adaptive processes provide negative feedback, generating localized activity patterns. Synaptic connectivity in our model is described by an integral kernel that weakens dynamically due to activity-dependent synaptic depression, leading to marginally stable traveling fronts (with attenuated backs) or pulses of a fixed speed. Our analysis quantifies how weak stimuli shift the relative position of these waves over time, characterized by a wave response function we obtain perturbatively. Persistent and continuously visible stimuli model moving visual objects. Intermittent flashes that hop across visual space can produce the experience of smooth apparent visual motion. Entrainment of waves to both kinds of moving stimuli are well characterized by our theory and numerical simulations, providing a mechanistic description of the perception of visual motion.}

\keywords{visual object motion, neural field, traveling waves, synaptic depression}

\maketitle

\section{Introduction}
\label{intro}

Coherent neural activity patterns respond to and even predict sensory stimuli~\citep{ERMENTROUT2001}. The spatiotemporal dynamics that emerge in the context of sensory processing can be substantially complex but nevertheless reproducible, implying internal features of neural populations organize activity responses in repeatable ways~\citep{WU2008}. Along these lines, the careful characterization of these evoked dynamics across trials can provide insight into network structure and its role~\citep{XU2007}. Moreover, repeatedly evoked coherent patterns of activity can reverberate even in the spontaneous dynamics that follow stimulus trials~\citep{HAN2008}. These results suggest that the coherent spatiotemporal activity dynamics that emerge in sensory cortices following stimulus presentations subserve computations determining animals' future expectations and behavior~\citep{ZANOS2015}.

Visual cortical waves are a well studied example of coherent cortical dynamics, which are generated both by electrical and visual stimulation~\citep{WU2008}. The visual system continually converts ongoing and complex input into abstract but appropriately detailed representations~\citep{TENENBAUM2011}. These computations serve to not only represent position, motion, and shape of objects~\citep{BORN2005}, but also to resolve ambiguities including anticipated changes~\citep{KNILL2004}. As this process unfolds over time, new observations are merged with previous estimates, potentially inferring object features via the spatiotemporal neural activity of corresponding cortical networks~\citep{CICHY2014,BILL2022}. Moreover, propagating neural activity waves resulting from complex visual stimuli, even those arising in visual cortex, appear to subserve the onset and specifics of motor outputs, like saccades~\citep{ZANOS2015}.

Object motion tracking and prediction are important visual functions for animals behaving in their natural environments~\citep{ECKERT2001}. Flying predators must accurately track and predict the movement of prey animals along the ground to plan their pursuits and strikes~\citep{KANE2014}, while flocking and herding animals anticipate and rapidly respond to the movements of their neighbors to avoid collisions and stay together~\citep{NAGY2010,TORNEY2018}. Often, conspecifics or objects are only intermittently seen by animals, requiring their velocity and position estimates to be made during the occlusion periods~\citep{OREILLY2008}. Such abilities suggest a normative account of the {\em apparent motion illusion} in which successive stationary flashes at different locations are perceived as a single moving object hidden from view between flashes~\citep{RAMACHANDRAN1986}. Voltage sensitive dye recordings from awake fixating monkeys observing such stimuli reveal an interaction of neural activity waves with external inputs suggesting early visual cortical activity helps represent perception of a moving object and its velocity~\citep{CHEMLA2019}. This study also proposed a detailed computational model in which a suppressive wave of activity is generated by the second of two flashes, either explaining away the ambiguity of the first flash as possibly another object or representing the two flashes as a single moving object.

Here, we analyze a neural field description of the apparent motion illusion that relies on the entrainment of traveling activity wave solutions to a sequence of transient and localized stimulations. Neural fields model neuronal networks as a continuous and spatially-extended excitable medium described by nonlinear integrodifferential equations allowing for direct analysis using methods adapted from nonlinear partial differential equations, such as reaction diffusion models~\citep{BRESSLOFF2012}. Such a framework is ideal as it affords analytical treatments of the local network mechanisms underlying emergent spatiotemporal patterned activity~\citep{HUANG2004,GOULET2011} and stimulus-behavior relationships commonly recorded in cognitive tasks~\citep{BRESSLOFF2012B,KILPATRICK2018,ERLHAGEN2006}. Traveling wave solutions can be identified explicitly in many instances~\citep{PINTO2001,COOMBES2005}, as can their response to non-trivial stimuli~\citep{FOLIAS2005,ERMENTROUT2010,AMARI1977}. Our model incorporates a physiologically realistic form of negative feedback as short term synaptic depression, leading to an attenuation at the back of waves, producing traveling pulses~\citep{KILPATRICK2010A}. We will solve for traveling waves and identify their response to flashing stimuli, interpreting the resulting dynamics as a stimulus motion percept. Such an approach allows for explicit and dynamical characterization of the conditions required to promote the apparent motion illusion across a range of potential stimulus types.

Our perturbative approach to studying how traveling waves respond to transient or weak stimuli specifically estimates how a wave's position changes in response to inputs. Linear asymptotics and even weakly nonlinear analysis have been used previously to understand how perturbations in synaptic connectivity, input, or model parameters shape waves and patterns in neural field equations~\citep{BRESSLOFF2003,VENKOV2007,BRESSLOFF2001,COOMBES2005,AMARI1977}. Perturbative theories describing how waves transform inputs and synaptic weight heterogeneities into changes in position and speed have been used as a model of idiothetic position (i.e., where an animal is or what direction they are heading)~\citep{ZHANG1996,XIE2002,BURAK2009}. Weak inputs alter the dynamics enough to displace wave positions without substantially disrupting their shape, allowing for an accurate linear input-response theory~\citep{KILPATRICK2012}. Since our model incorporates nonlinear negative feedback, care must be taken in performing the asymptotic calculations to characterize the response to inputs~\citep{KILPATRICK2010C}. In addition, we can study the differential effects of inputs to the synaptic depression variable, which recovers more slowly, as opposed to the neural activity variable.

The response of traveling waves to transient inputs can be accurately captured by our perturbation theory, providing closed form expressions for the distance traveling waves are shifted by perturbations. Realizing wave position as a model of inferred object motion, we conclude that these position shifts encode the history of encountered inputs. Flash sequences that hop to new locations according to effective speeds that are sufficiently close to the intrinsic speed of a traveling wave can entrain it. Our framework precisely characterizes the width of this band of entrainable speeds as a function of the amplitude and frequency of the input, providing testable predictions concerning the psychophysics of apparent motion.

\section{Neural field model with synaptic depression}\label{sec:model}

Prior models of local negative feedback in neural fields often employ heuristics like linear adaptation~\citep{PINTO2001}, not based directly on physiology. Continual activation of neurons can transiently reduce the efficacy of synapses originating from them, often due to vesicle depletion~\citep{FORTUNE2001}. Activity-based models of neural activity, like neural fields, typically incorporate such short-term depression according to the Tsodyks-Markram model, derived by temporally smoothing the dynamically evolving efficacy of resource-dependent synapses~\citep{TSODYKS1998,BART2005}. Dynamic reductions in the strength of synaptic weights originating from recently active neural populations curtails spatiotemporal activity and can produce propagating waves and patterns in continuum neural fields~\citep{YORK2009,KILPATRICK2010B,BRESSLOFF2012B}. We will move beyond these prior studies to consider the effects of transient and persistent inputs upon waves in networks subject to synaptic depression, and consider how such a model can represent visual motion encoding. Specifically, we consider the following integro-differential equation system:
\begin{subequations} \label{eqn:model}
\begin{align}
	\frac{\partial}{\partial t} u(x, t) &= 
        -u + w \ast (q f[u]) + \ep I_u(x,t), \label{eqn:model_u} \\
	\tau_q \frac{\partial}{\partial t} q(x,t) &= 
        1 - q - \beta q f[u(x,t)] + \ep I_q(x,t). \label{eqn:model_q}
\end{align}
\end{subequations}
Here, $u(x,t)$ and $q(x,t)$ denote the average normalized voltage and synaptic efficacy, respectively, at location $x$ and time $t$, and $f(u)$ represents the output firing rate~\citep{ERMENTROUT1998}.
The synaptic timescale $\tau_q$ is generally chosen to be longer than the non-dimensionalized time units $\tau_q > 1$ representing the timescale of neural activity~\citep{TSODYKS1998}.
Both $u$ and $q$ are normalized so that typical activity (in the absence of stimuli, and with suitable initial conditions) places each variable in the interval $[0, 1]$ for any particular point in space-time. 
The weight kernel $w(x,y) \equiv w(|x-y|)$ is purely distance-dependent and gives the density of synaptic connections from pre-synaptic neurons at location $y$ to post-synaptic neurons at location $x$, as is common~\citep{WILSON1973, AMARI1977,PINTO2001} (See \cite{BRESSLOFF2001,KILPATRICK2008} for examples of the effects of breaking this translation symmetry). 
The non-local spatial operator takes the form of a convolution $w \ast (qf[u]) = \int_{\mathbb{R}} w(x-y) q(y,t) f[u(y,t)] dy$. 
We will take the weight kernel to be a decaying exponential $w(x) = \tfrac{1}{2}e^{-|x|}$ for explicit calculations.

The firing-rate function $f$ is non-linear, monotonic, and is often normalized so it is lower bounded to 0 for small values and saturates to 1 for large values. 
Common choices are sigmoidal or Heaviside functions. 
Throughout this paper, we will choose $f(u) = H(u - \theta)$ where $H$ is the Heaviside function, and $\theta$ is a threshold parameter describing the input activity needed to generate output activity from a local population. 
The binary output of the Heaviside function approximates firing rates to simply be {\em low} or {\em high} but allows us to, for any given time $t$, partition the spatial domain into an active region $\{x \in \RR \mid f(u(x,t)) = 1\}$, inactive region $\{x \in \RR \mid f(u(x,t)) = 0\}$ and a set of threshold crossings $\{x \in \RR \mid u(x,t) = \theta\}$~\citep{COOMBES2012}. 
Spatiotemporal inputs $I_u(x,t)$ and $I_q(x,t)$ representing the impact of visual stimuli are assumed to be weak ($\ep \ll 1$) compared to ongoing activity and may arise in either the neural activity or synaptic efficacy variables.

The strength of synaptic depression is parameterized by the rate $\beta > 0$~\citep{TSODYKS1998,BART2005}.
Notice that in regions where $u>0$, we have $f[u] = 1$, and then Eq.~(\ref{eqn:model_q}) can be written $\gamma \tau_q q_t = \gamma - q + \gamma \varepsilon I_q$, where $\gamma = \frac{1}{1+\beta} \in (0, 1]$. 
Under this formulation, $\tau_q$ is the timescale of synaptic replenishment, $\gamma\tau_q$ is the timescale of synaptic depletion, and $\gamma$ is the ratio between them. 
For $\gamma \approx 1$ ($0< \beta \ll 1$), the timescales of replenishment and depression are similar, and when $\beta$ is not small, the depression timescale is shorter and thus depression happens at a faster rate. 
Inclusion of synaptic depression dynamically reduces synaptic efficacy ($q < 1$), attenuating the active regions within waves. 
In the next section, we characterize front and pulse solutions more precisely through explicit construction.

\section{Traveling Wave Solutions}
The neural field model with synaptic depression supports a variety of traveling wave solutions that are qualitatively different than the corresponding scalar model~\citep{KILPATRICK2010B} without synaptic depression.
In a purely excitatory scalar neural field model,
\begin{align}
    \frac{\partial}{\partial t}u(x,t) = - u + w*f[u], \label{scalar}
\end{align}
a sufficiently large, initially active region will spread indefinitely as counterpropagating traveling fronts in the long time limit~\citep{FAYE2018}. Traveling front solutions, which take the form of heteroclinics connecting the active ($u = 1$) to the inactive ($u=0$) state in traveling wave coordinates $\xi = x- ct$ can be constructed explicitly in the case of step nonlinearities $f(u) = H(u-\theta)$~\citep{PINTO2001} and using a homotopy argument in the case of smooth nonlinearities~\citep{ERMENTROUT1993}. That is, for $u \equiv U(\xi)$, we have a solution that satisfies $\lim_{\xi \to - \infty} U(\xi) = 1$ and $\lim_{\xi \to + \infty} U(\xi) = 0$. Our analysis of traveling waves in the model with synaptic depression hinges upon the manipulation of the left limit.

The different form of traveling waves in the model with synaptic depression emerges due to attenuation of the activity level within the active region.
To demonstrate this, we examine the fixed points of the analogous {\em space-clamped} equations of Eq.~(\ref{eqn:model}) in the absence of inputs ($I_u \equiv I_q \equiv 0$) obtained by assuming $u(x,t) \equiv u(t)$ and $q(x,t) \equiv q(t)$, so $ u'(t) = -u + qf[u]$ and $\tau_q q'(t) = 1-q - \beta q f[u]$, where we have assumed the normalization $\int_{\mathbb{R}} w(x) dx = 1$.
For $f[u] = H(u-\theta)$, we always have the quiescent fixed point $(\bar{u},\bar{q}) = (0,1)$ and if $\gamma > \theta$ we will also have $(\bar{u}, \bar{q}) = (\gamma, \gamma)$.
As we will show, if $\gamma > \theta$, traveling front solutions of Eq.~(\ref{eqn:model}) for $I_u \equiv I_q \equiv 0$ can be supported with an attenuated active state, but if $\gamma < \theta$, traveling pulse solutions with a finite active region can emerge.

Previously, in \cite{KILPATRICK2010B}, traveling front and pulse solutions were characterised for a generalization to the model Eq.~(\ref{eqn:model}) that incorporated spike rate adaptation.
Here we rederive and extend some of these results (See Figs.~\ref{fig1:fronts} and \ref{fig:pulse_bifurcation}) in anticipation of our later derivations. Analyzing the model without spike frequency adaptation allows us to more clearly characterize the impact of synaptic depression.
We also identify a previously overlooked traveling wave solution, the stable retreating front, which co-exists with the stable advancing front and emerges due to the negative feedback.

\subsection{Fronts}

Changing variables to wave coordinates $\xi = x - ct$ in Eq.~(\ref{eqn:model}) and assuming $u \equiv U(\xi)$ and $q \equiv Q(\xi)$, we find that such solutions must satisfy
\begin{align*}
    -c U'(\xi) &= -U(\xi) + \int_{\mathbb{R}} w(\xi - y) Q(y) f(U(y)) dy, \\
    -c \tau_q Q'(\xi) &= 1 - Q(\xi) - \beta Q(\xi) f(U(\xi)).
\end{align*} 
Traveling fronts in an excitatory neural field (with $w(x) >0$) with a step nonlinearity $f(u) = H(u - \theta)$ can be shifted such that the single threshold crossing point occurs at $\xi = 0$, so that $U(0) = \theta$. The active region is thus $\{ \xi \in \RR \ | \ U(\xi) > \theta \} = ( - \infty, 0)$. Note, that the function $U(\xi)$ is not necessarily monotone decreasing in $\xi$ as it is in the scalar system due to the negative feedback resulting from synaptic depression. Thus, we must impose an additional inequality $U(\xi) > \theta$ for $\xi < 0$, which should be checked for self-consistency: 
\begin{subequations} \label{fronteqns}
\begin{align}
    -c U'(\xi) &= -U(\xi) + \int_{- \infty}^0 w(\xi - y) Q(y) dy, \label{eqn:fronts_U} \\
    -c \tau_q Q'(\xi) &= 1 - Q(\xi) - \beta Q(\xi) H(-\xi). \label{eqn:fronts_Q}
\end{align} 
\end{subequations}
Eq.~(\ref{eqn:fronts_Q}) is now piecewise linear, decoupled from Eq.~(\ref{eqn:fronts_U}), and can be solved piecewise along with enforcing continuity and appropriate boundary conditions depending on the direction (${\rm sign} (c)$) of travel. Before analyzing the case of moving fronts ($c \neq 0$), we examine the degenerate case of a standing wave solution with speed $c = 0$. In this case, Eq.~(\ref{eqn:fronts_Q}) reduces to a stationary equation, we have $\xi \equiv x$, and the profile of synaptic efficacy is given by
\begin{align*}
    Q(x) = \left\{ \begin{array}{cc} 1, & \ x>0, \\ \frac{1}{1+\beta} \equiv \gamma, & \ x<0. \end{array} \right.
\end{align*}
Substituting into Eq.~(\ref{eqn:fronts_U}), we see
\begin{align*}
    U(x) = \gamma \int_{- \infty}^0 w(x-y) dy,
\end{align*}
so for a normalized ($\int_{\mathbb{R}} w(x) dx = 1$) and even ($w(-x) = w(x)$) weight function, we have $\theta = U(0) = \frac{\gamma}{2}$, implying standing fronts only arise for a specific choice of the depression rate $\beta = \frac{1}{2\theta} - 1$ which perfectly balances the tendency of the active region to invade inactive regions with the rate of activity decay~\citep{ERMENTROUT1993}.

\begin{figure}[t!]
\begin{center} \includegraphics[width=17cm]{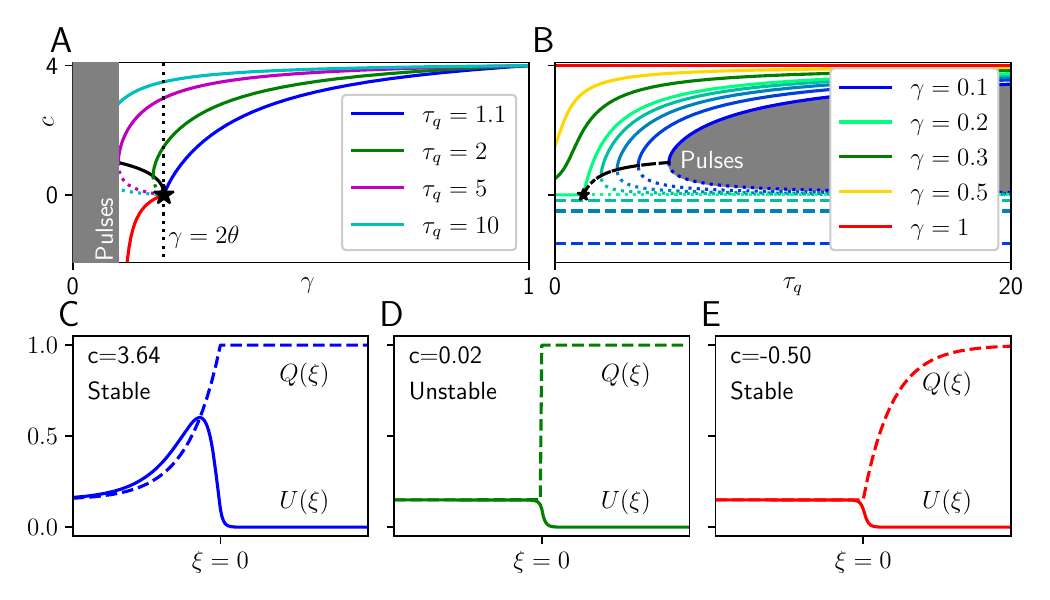} \end{center}
\caption{
{\bf Traveling fronts in a neural field with synaptic depression.}
{\bf A.} Speeds of stable (solid) and unstable (dashed) traveling fronts as a function of the $\gamma = \frac{1}{1+\beta}$ (synaptic depletion/replenishment timescale ratio).
For $\gamma < \theta$, self-consistency is broken, and traveling pulses emerge (see Section \ref{sec:pulses}), and for $\theta < \gamma < 2\theta$, retreating front solutions emerge (red), with speed and profile independent of $\tau_q$. 
Stable speed $c$ curves coalesce as $\gamma \to 1$ as effects of synaptic depression vanish.
{\bf B.} Speeds as a function of synaptic timescale $\tau_q$, clearly showing that the speed of retreating fronts is independent of $\tau_q$. Black line indicates saddle node bifurcation where the stable and unstable advancing front speeds meet.
{\bf C, D, E.} Example profiles of fast stable advancing (blue), slow unstable advancing (green), and retreating (red) fronts. 
Note the $\xi$-dependent region of $Q(\xi)$ trades places between advancing and retreating cases.
For all panels $\theta = 0.1$; for the profile panels $\gamma = 0.15$ and $\tau_q = 20$.
}
\label{fig1:fronts}
\end{figure}

For forward moving fronts ($c >0$), we solve Eq.~(\ref{eqn:fronts_Q}) with the boundary conditions $\lim_{\xi \to \infty}Q(\xi) = 1$ and $\lim_{\xi \to - \infty} Q(\xi) = \gamma$, yielding
\begin{align*}
    Q(\xi) = \begin{cases}
        1, & \ \xi >0, \\ 
        \gamma + (1- \gamma) \exp \left( \frac{\xi}{c\gamma \tau_q} \right) , & \ \xi < 0.
    \end{cases}
\end{align*}
Substituting this into Eq.~(\ref{eqn:fronts_U}) and solving  the $\xi > 0$ case with integrating factors gives 
\begin{align*}
    U(\xi) &= e^{\xi/c} \left( U_0 - \frac{\gamma + c\gamma \tau_q}{2(1+c)(1+c\gamma\tau_q)} \right) + e^{-\xi} \frac{\gamma + c\gamma \tau_q}{2(1+c)(1+c\gamma\tau_q)},
\end{align*}
which we bound as $\xi \to \infty$ by requiring the coefficient on $e^{\xi/c}$ to be zero. Recalling that $U(0) = \theta$ we then have
\begin{align}
    \theta = \frac{\gamma + c\gamma \tau_q}{2(1+c)(1+c\gamma\tau_q)} \iff (2\theta \gamma \tau_q)c^2 + (2\theta+2\theta\gamma\tau_q - \gamma\tau_q)c + (2\theta-\gamma) = 0, \label{eqn:frontspeedpos}
\end{align}
whose solutions $c$ are roots of a quadratic:
\begin{align*}
    c_{\pm} = \frac{\gamma \tau_q - 2 \theta (1 + \gamma \tau_q) \pm \sqrt{4 \theta^2 (1- \gamma \tau_q)^2 -  4 \theta \gamma \tau_q (1 + \gamma (\tau_q-2)) + \gamma^2 \tau_q^2}}{4 \theta \gamma \tau_q},
\end{align*}
plotted in Fig.~\ref{fig1:fronts}A,B. We can also solve the $\xi < 0$ case, which automatically satisfies our boundedness condition, and completes the profile for advancing fronts
\begin{align*}
    U(\xi) &= \begin{cases}
        \theta e^{-\xi}, & \ \xi > 0, \\
        \gamma + K_1 e^{\xi/(c\gamma\tau_q)} + K_2 e^{\xi} + \left(\theta - \gamma - K_1 - K_2\right) e^{\xi/c} , & \ \xi < 0, 
    \end{cases} \\[3ex]
    K_1 &= \frac{c\gamma\tau_q}{1 - \gamma\tau_q} \cdot \frac{(1-\gamma)c\gamma^2\tau_q^2}{1-(c\gamma\tau_q)^2}, \ \ \ \ \ \ \ \ K_2 = -\frac{1}{2(1-c)} \cdot \left( \frac{(1-\gamma) c \gamma \tau_q}{1 - c \gamma \tau_q} - \gamma \right).
\end{align*}
The self-consistency inequality ensuring the active region remains superthreshold then requires $\lim\limits_{\xi \to - \infty} U(\xi) = \gamma > \theta$ or $1/(1 + \beta) > \theta$ implying $\beta < (1 - \theta)/\theta$. Profiles are plotted in Fig.~\ref{fig1:fronts}C,D.

On the other hand, if we assume $c<0$ (retreating fronts), we can similarly find solutions for which the quiescent region invades the active region. We again solve Eq.~(\ref{eqn:fronts_Q}), with the same boundary conditions, but now having $c<0$ implies
\begin{align*}
    Q(\xi) &= \begin{cases} 
        1 + (\gamma-1)e^{\xi/(c\tau_q)}, & \ \xi > 0, \\
        \gamma, & \ \xi < 0.
    \end{cases}
\end{align*}
Again we substitute into Eq.~(\ref{eqn:fronts_U}), enforce boundedness and the threshold condition $U(0) = \theta$. After integrating, we find a condition determining the speed of the front
\begin{align*}
    c = \frac{\gamma - 2\theta}{2\gamma - 2\theta}, 
\end{align*}
and a formula for the activity variable of form
\begin{align} \label{regressfront}
    U(\xi) = \begin{cases}
        \theta e^{-\xi}, & \xi > 0, \\
        \gamma + (\theta - \gamma) e^\xi, & \xi < 0.
    \end{cases}
\end{align}
Speeds are plotted in Fig.~\ref{fig1:fronts}A,B and example profile in Fig.~\ref{fig1:fronts}E. Such fronts exist for sufficiently strong depression as bounded by the standing front condition $\beta > (1- 2 \theta)/(2\theta)$ or $\gamma < 2 \theta$, but not too strong ($\beta < (1-\theta)/\theta$ or $\gamma > \theta$).

The speeds of stable progressing traveling waves decrease with the amplitude $\beta$ of synaptic depression (increasing with $\gamma = 1/(1+ \beta)$, the ratio of synaptic depletion and replenishment timescales, Fig.\ref{fig1:fronts}A). The parameter $\gamma$ defines the steady state average voltage in the active region so fronts cannot exist if $\gamma < \theta$, and for $\theta < \gamma < 2\theta$, a retreating traveling front emerges (red curve). A branch of unstable and slow traveling fronts also exists in this region. Complementarily, wavespeeds increase with the 
synaptic depression timescale $\tau_q$ (Fig.~\ref{fig1:fronts}B). Retreating fronts (Fig.~\ref{fig1:fronts}E) are not simply reflections of advancing fronts (Fig.~\ref{fig1:fronts}C,D). Rather, an advancing (retreating) front evolves as the active (inactive) region invades the inactive (active) region, so the synaptic depression is constant in the inactive (active) region. In contrast, traveling pulses are always advancing, so that a finite length active regions moves into an inactive region, but can be reflected to produce a solution that moves the opposite direction.

\subsection{Pulses} \label{sec:pulses}
Sufficiently strong synaptic depression produces traveling pulses as negative feedback brings the average voltage $u$ below threshold $\theta$ to create a pulse back (if $\theta > \gamma = \frac{1}{1+\beta}$). These solutions are characterized by both a speed $c$ and width $\Delta$ which are determined by the model parameters.

\begin{figure}[t!]
\begin{center} \includegraphics[width=16cm]{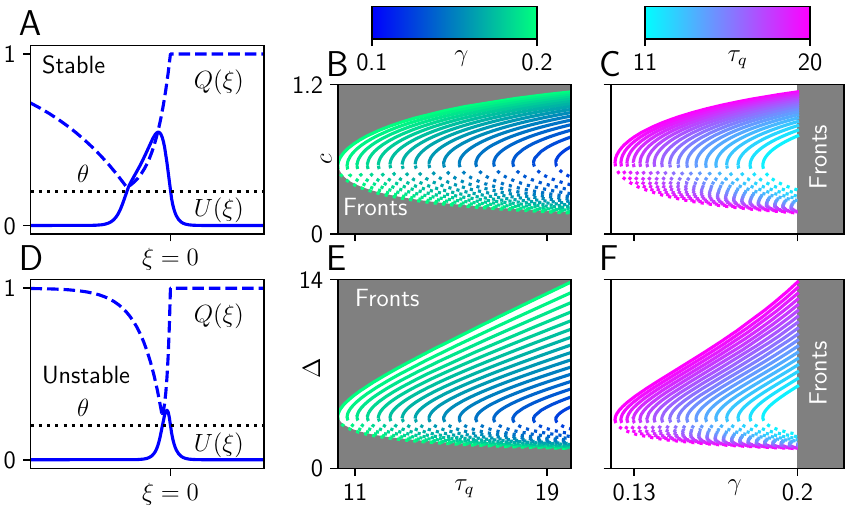} \end{center}
\vspace{2mm}
\caption{
    \textbf{Traveling pulses in a neural field with synaptic depression.}
    (\textbf{A,D})
    Stable and unstable pulse profiles for parameters $\tau_q = 20$, $\gamma = \tfrac{1}{6}$, (or equivalently $\beta = 5$), and $\theta = 0.2$.
    (\textbf{B,E}) 
    The pulse speed $c$ and width $\Delta$ are plotted as a function of the synaptic timescale $\tau_q$ with color indicating $\gamma$, solid lines indicating stable branches and dotted lines indicating unstable branches.
    (\textbf{C,F})
    The pulse speed $c$ and width $\Delta$ as a function of $\gamma$ with color indicating $\tau_q$, solid lines indicating stable branches and dotted lines indicating unstable branches.
    We see that as the synaptic efficacy time-scale becomes shorter ($\tau_q$ small), depression is more rapid and both the width and speed of the pulse shrink.
    Similarly increasing the strength of depression ($\beta$ large) or equivalently shortening the effective timescale of synaptic depression ($\gamma$ small) will also reduce the width and speed.
}
\label{fig:pulse_bifurcation}
\end{figure}

Changing variables to traveling wave coordinates and considering step nonlinearity $f(u) = H(u - \theta)$, we define an active region $\{ \xi \in \mathbb{R} \ | \ U(\xi) > \theta \} = (- \Delta, 0)$ so the pulse crosses threshold twice at $\theta = U(0) = U(-\Delta)$. In the absence of stimulus, the system Eq.~(\ref{eqn:model}) then becomes
\begin{subequations} \label{pulseeqns}
\begin{align}
    -cU'(\xi) &= - U(\xi) + \int_{- \Delta}^0 w(\xi - y) Q(y) dy, \label{eqn:pulses_U} \\
    -c \tau_q Q'(\xi) &= 1 - Q(\xi) - \beta Q(\xi) \left[ H(\xi + \Delta) - H(\xi) \right]. \label{eqn:pulses_Q}
\end{align}
\end{subequations}
Without loss of generality, we take $c>0$, since backward moving pulses are simply reflections of forward moving pulses. The synaptic efficacy equation is now a decoupled piecewise linear equation. Enforcing the boundary conditions $\lim\limits_{\xi \to \pm \infty} Q(\xi) = 1$, we obtain the solution
\begin{align*}
    Q(\xi) = \begin{cases}
        1, & 0 \le \xi, \\
        \gamma + (1-\gamma) e^{\xi/(c \gamma \tau_q)}, &  -\Delta \le \xi < 0, \\
        1 - \left[1 - Q(-\Delta)\right] e^{(\xi + \Delta)/(c\tau_q)} , & \xi < - \Delta.
    \end{cases}
\end{align*}
Substituting back into Eq.~(\ref{eqn:pulses_Q}), we find
\begin{align*}
    -c U'(\xi) = -U(\xi) + \frac{1}{1+ \beta} \int_{- \Delta}^0 (1 + \beta e^{(1 + \beta) y/(c \tau_q)}) w(\xi - y) d y,
\end{align*}
which can be solved piecewise up to free constants that can be identified by enforcing continuity, boundedness, and boundary conditions. Notice in the limit $\Delta \to \infty$, we recover the equation for $U(\xi)$ in the front case. For the exponential weight function, and the threshold conditions $U(- \Delta) = U(0) = \theta$ we can then obtain the following equations for the wavespeed $c$ and pulsewidth:
\begin{subequations} \label{pulsewidspeed}
\begin{align}
    \theta =& \frac{c \gamma \tau_q}{2 (c+1)(c \gamma \tau_q + 1)} \left[ 1 - \gamma e^{- \Delta} \left[ 1 + e^{- \Delta/(c \gamma \tau_q)}\right]\right],\\
    \theta =& \frac{(2c+1) \gamma}{2(c+1)} \left( 1- e^{-\Delta/c} \right)  + \frac{\gamma}{2(c-1)} \left[ 1 + \frac{c (1- \gamma) \tau_q}{c \gamma \tau_q -1 } \right] \left( e^{- \Delta} - e^{- \Delta/c} \right) \\
    & - \frac{c (1- \gamma) \tau_q e^{- \Delta / (c \gamma \tau_q)}}{2 (c+1)(c \gamma \tau_q + 1)}  + \frac{c^2 \gamma^3 (1- \gamma) \tau_q^3 (e^{- \Delta/(c \gamma \tau_q)} - e^{- \Delta/c})}{(c^2 \gamma^2 \tau_q^2 - 1)(\gamma \tau_q - 1)}. \nonumber
\end{align}
\end{subequations}
Numerical root-finding techniques can subsequently be used to identify the speed $c$ and width $\Delta$ of traveling pulse solutions given implicitly by Eq.~(\ref{pulsewidspeed}).
For sufficiently weak synaptic depression (corresponding to small $\beta$ and large $\gamma$), there are coexisting branches of (stable) wide and (unstable) narrow traveling pulse solutions (Fig.~\ref{fig:pulse_bifurcation}). Increasing either $\gamma$ or $\tau_q$ increases the speed and width of the stable traveling pulses. Explicit results on linear stability could be obtained by extending the weak methods developed for piecewise smooth pseudolinear operators developed in \cite{KILPATRICK2010C}, but here we have determined branch stability by full numerical simulation. With the traveling pulse solutions in hand, and noting they are generally sufficiently wide so interactions between the front and back are weak ($0< e^{- \Delta} \ll 1$), we now proceed with performing an asymptotic analysis determining how weak forcing modeling sensory input guides the position of a pulse over time.

\section{Wave Response Function} \label{sec:wave_resposne}

Prior studies of the impact of external inputs on spatiotemporal patterns in neural fields have both considered cases in which direct construction is possible as well as perturbative studies in which the impact of forcing is taken to be weak. 
Special cases which allow for the calculation of explicit solutions, have revealed the onset of complex dynamics using linear stability or even weakly nonlinear analysis yielding descriptions of breathers~\citep{FOLIAS2004,FOLIAS2005,FOLIAS2011} or even oscillations reminiscent of perceptual rivalry~\citep{LOXLEY2009,KILPATRICK2010D,BRESSLOFF2012B}. 
Numerical studies have also been used to identify the emergence of topological defects in patterns based on transient and localized inputs~\citep{HUTT2003}. 
However, in cases of weak forcing, it is possible to follow the asymptotic approach of \cite{AMARI1977}, assuming waves or bumps approximately retain their shape but are shifted in space due to the projection of the input along the pattern's marginally stable direction~\citep{ZHANG1996,BENYISHAI1997,WU2008B,BURAK2009,ERMENTROUT2010,ITSKOV2011}.   
Effects of a weak but otherwise arbitrary spatiotemporal input can be approximated by deriving a pattern's corresponding spatiotemporal filter determined by the adjoint of the linearization about the pattern~\citep{ERMENTROUT2010,KILPATRICK2012}.
Such an approach bears resemblance to theory describing how perturbations phase shift nonlinear oscillators, which defines a phase response function~\citep{ERMENTROUT1996,BROWN2004}, so we refer to our formulation as the {\em wave response function}.
We consider the impact of a nonlinear auxiliary variable in the neural field (synaptic depression) on such input-response relationships, defining an entrainment problem for a set of periodic spatiotemporal inputs, as a means of framing visual motion processing. 
Ultimately, this allows us to describe potential neural mechanisms to account for behaviorally reported visual phenomena like the apparent motion illusion.

\subsection{General Framework}

As shown above, in the absence of inputs ($I_u(x,t) \equiv I_q(x,t) \equiv 0$), Eq.~(\ref{eqn:model}) can support traveling wave solutions $(U(\xi), Q(\xi))$ that can be determined explicitly for step nonlinearities $f(u) = H(u-\theta)$ and weight kernels with defined integrals.
The speed $c$ (and for pulses, the width $\Delta$) can be specified implicitly and thus calculated with root finding. 
Weak stimulation perturbatively shapes both the profile and position (within the original traveling coordinate frame) of these waves. However, importantly, waves are linearly stable to shape-changing perturbations but marginally stable to those that shift their position within the wave coordinate frame~\citep{AMARI1977,PINTO2001,COOMBES2004}. Since we treat wave positions as an internal model of the represention of visual object location, we are concerned with wave position shifts as they determine entrainment to moving stimuli of different speeds. To derive an approximate equation for the time-dependent evolution of the wave position $\ep \nu (t)$ relative to the coordinate frame $\xi = x - ct$, we begin by changing variables in Eq.~(\ref{eqn:model}) from $(x,t)$ to $(\xi,t)$. As a result, we have an evolution equation for $u(\xi,t)$ and $q(\xi,t)$ defined
\begin{subequations} \label{xiteqn}
\begin{align}
    -c u_{\xi} + u_t &= -u + w*(qf[u])+\ep I_u(\xi,t), \\
    -c \tau_q q_{\xi} + \tau_q q_t &= 1-q - \beta q f[u] + \ep I_q(\xi,t),
\end{align}
\end{subequations}
where Eq.~(\ref{xiteqn}) generalizes solutions in the traveling frame to allow them to be time-dependent. Effects of external perturbations representing sensory stimuli are then accounted by considering the perturbation expansion
\begin{subequations} \label{pertexp}
\begin{align}
    u(\xi,t) &= U(\xi - \ep \nu (t)) + \ep \phi(\xi - \ep \nu (t),t) + \cdots, \\
    q(\xi,t) &= Q(\xi - \ep \nu (t)) + \ep \psi (\xi - \ep \nu (t),t) + \cdots,
\end{align}
\end{subequations}
where $(U(\xi) ,Q (\xi))$ is an unperturbed wave solution and $\ep \nu(t)$ is a small time-dependent function determining position shifts in the traveling coordinates generated by the inputs. The terms $\phi (\xi) $ and $\psi (\xi) $ weakly modify the wave shape. Plugging the expansion Eq.~(\ref{pertexp}) into Eq.~(\ref{xiteqn}), we find the ${\mc O}(\ep)$ terms satisfy the inhomogeneous linear system
\begin{align}
    \left( \begin{array}{c} \phi_t \\ \tau_q \psi_t \end{array} \right) + \LL \left( \begin{array}{c} \phi \\ \psi \end{array} \right) = \nu'(t) \left( \begin{array}{c} U'(\xi) \\ \tau_q Q'(\xi) \end{array} \right) + \left( \begin{array}{c} I_u(\xi + \ep \nu,t) \\ I_q(\xi + \ep \nu, t) \end{array} \right), \label{eqn:first_order}
\end{align}
where $\LL$ is the linear operator
\begin{align*}
    \LL \left( \begin{array}{c} \phi \\ \psi \end{array} \right) = -
    c\begin{pmatrix}
        \phi_\xi \\ \tau_q \psi_\xi
    \end{pmatrix}
    +
    \begin{pmatrix}
        \phi \\ \psi 
    \end{pmatrix}
    +
    \begin{pmatrix}
        - w*\left[ Q f'(U) \phi + \psi f(U)\right] \\
        \beta Qf'(U)\phi + \beta f(U) \psi
    \end{pmatrix}.
\end{align*}
A bounded solution of Eq.~(\ref{eqn:first_order}) exists if its right hand side is orthogonal to all elements of the null space of the adjoint operator $\LL^*$~\citep{KEENER2000A,KEENER2000B,BRESSLOFF2001}. Defining the $L^2$ inner product $\langle \vecf, \vecv \rangle = \int_{\RR} \mathbf{g}^T(\xi) \mathbf{f}(\xi) d \xi  $, we can then identify the adjoint operator by requiring $\langle {\mc L} \vecu, \vecv \rangle = \langle \vecu, {\mc L}^* \vecv \rangle$. Integrating by parts and rearranging integrals in this equation, we obtain the definition
\begin{align}
\LL^* \left( \begin{array}{c} \phi \\ \psi \end{array} \right) =
    c\begin{pmatrix}
        \phi_\xi \\ \tau_q \psi_\xi
    \end{pmatrix}
    +
    \begin{pmatrix}
        \phi \\ \psi 
    \end{pmatrix}
    +
    \begin{pmatrix}
        - Q f'(U) \cdot w \ast \phi + \beta Q f'(U) \psi \\
        - f(U) \cdot w \ast \phi + \beta f(U) \psi
    \end{pmatrix},
\end{align}
so the one-dimensional null space $(v, p)^T$ defined $\LL^*(v, p)^T = (0,0)^T$ satisfies the system
\begin{subequations}
\begin{align}
    c v_{\xi} &= - v + f'(U) Q \cdot w*v - \beta Q f'(U) p, \\
    c \tau_q p_{\xi} &= -p + f(U) \cdot w*v - \beta f(U) p.
\end{align}
\label{eqn:nullspace}
\end{subequations}
We apply our orthogonality condition to ensure boundedness and obtain the evolution equation
\begin{align}
\frac{d \nu}{dt} &= - \frac{\int_{\RR} \left[ v(\xi) I_u(\xi + \ep \nu, t) + p(\xi) I_q(\xi + \ep \nu,t) \right] d \xi}{\int_{\RR} U'(\xi) v(\xi) + \tau_q Q'(\xi) p(\xi) d \xi}, \label{eqn:asymptotic_ode}
\end{align}
revealing that the leading order description of wave perturbation arises through how the null vector filters inputs to the traveling wave. 

Our asymptotic approximation is thus determined implicitly as a solution to the nonlinear and non-autonomous differential equation, Eq.~(\ref{eqn:asymptotic_ode}). 
In the rest of this section, we analyze the predictions of this approximation and compare them to numerical simulations in response to transient inputs.
In the first case, we will study the impact of an abrupt and spatially homogeneous input upon the relative location of a propagating traveling front, showing the effects of synaptic depression. Second, we examine the effects of spatially localized inputs. 
Transient and abrupt inputs are assumed to shift in the relative position of the wave $\ep \nu (t)$ at a single localized moment in time, so the recurrence of Eq.~(\ref{eqn:asymptotic_ode}) can be ignored, allowing us to integrate the differential equation explicitly. Subsequently, we must treat the recurrence in Section \ref{sec:entrainment}, considering the effects of spatially localized but moving stimuli, since the $\nu$-dependence of the right hand side of Eq.~(\ref{eqn:asymptotic_ode}) is then non-trivial.

\subsection{Front Wave Response to a Global Flash}
\begin{figure}[t!]
\centering
\includegraphics[width=1\textwidth]{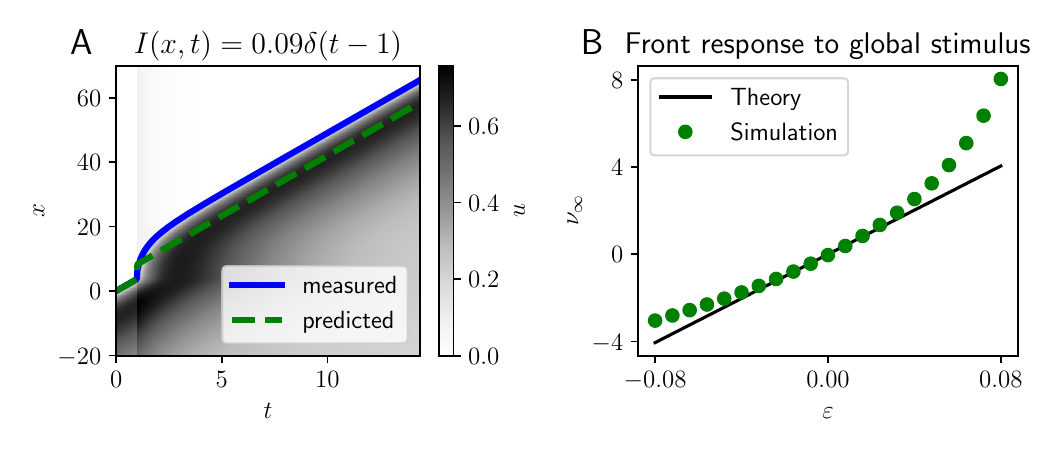}
\caption{
    {\bf Traveling front advanced by global flash stimulus.}
    {\bf A.} Spatiotemporal evolution of a traveling front perturbed by a spatially homogeneous, temporally pulsatile stimulus $\ep I_u(x,t) = \ep \delta (t-t_0)$ with $\ep = 0.09$. We compare the prediction (green dashed line) of our linear theory Eq.~(\ref{nuglob}) to the leading edge computed directly from simulation (blue line).
    {\bf B.} Wave response function $\nu_\infty$ predicted by the theory Eq.~(\ref{nuglob}) (black line) and compared to simulations (green dots). Weight function is exponential, firing rate nonlinearity is Heaviside, and model parameters are $\theta = 0.1$, $\gamma = 0.2$, and $\tau_q = 20$. 
}
\label{fig3}
\end{figure}

Here we consider the leading edge position of a traveling front as encoding a position, which is then perturbed by a spatially homogeneous and instantaneous in time input (Fig.~\ref{fig3}A). This basic case allows for explicit calculations which will guide our subsequent analysis of moving input tracking. We assume Eq.~(\ref{eqn:asymptotic_ode}) supports a traveling front solution and is stimulated only through the neural activity variable, $(I_u(\xi, t), I_q(\xi,t)) = (\delta(t - t_0), 0)$, arising via feed-forward input from upstream sensory areas. This could model a weak but broad visual light flash covering the receptive fields of a visual cortical region of interest. Recall that the perturbative scaling $\ep$ does not appear in Eq.~(\ref{eqn:asymptotic_ode}) as these are all factors themselves multiplied by $\ep$.

We first explicitly calculate the nullspace profile $(v (\xi), p (\xi))^T$ in the case of a traveling front solution whose adjoint linearization is Eq.~(\ref{eqn:nullspace}) with a Heaviside nonlinearity and so satisfying
\begin{align*}
    c \frac{dv}{d \xi} + v(\xi) &= \delta (\xi) \frac{Q(0)}{|U'(0)|} \int_{\RR}w(y)v(y) dy - \beta \delta (\xi) \frac{Q(0)}{|U'(0)|} p(0),  \\
    c\tau_q \frac{d p}{d \xi} + p(\xi) &= H(-\xi) \int_{\RR} w(\xi-y)v (y) dy - \beta H(-\xi) p(\xi).
\end{align*}
The Heaviside and delta distributions arise from the nonlinearity $f(U(\xi)) = H(U(\xi) - \theta) = H(- \xi)$ and its linearization $f'(U(\xi)) = H'(U(\xi) - \theta) = \delta(\xi)/ |U'(0)|$ which can be determined by noting
\begin{align*}
    - \delta(\xi) = \frac{d}{d \xi} H(- \xi) = \frac{d}{d\xi} H(U(\xi) - \theta) =  H'(U(\xi) - \theta) U'(\xi).
\end{align*}
Note the form of the $v(\xi)$ equation, $c v' (\xi) + v(\xi) = A \delta (\xi)$, suggests a bounded solution of the form $v(\xi) = H(\xi) e^{- \xi/c}$, whose step magnitude we are free to pick to be unity. No matter what scaling we pick for this term, it will be canceled in the fraction of Eq.~(\ref{eqn:asymptotic_ode}) since the terms of the null vector each appear once in the numerator and denominator.
Turning then to the $p(\xi)$ equation, we see that for $\xi > 0$, it takes the form $\frac{dp}{d \xi} = - \frac{1}{c \tau_q} p$ and 
$p(\xi) = B e^{-\xi/(c \tau_q)}$.
On the other half of the domain, $\xi < 0$, we substitute the form of $v(\xi)$ and compute the integral for an exponential weight function, finding
\begin{align}
    \frac{dp}{d \xi} + \frac{1}{c\gamma \tau_q} p &= \frac{1}{2(c+1)\tau_q} e^{\xi},  \label{peqxineg}
\end{align}
which can also be solved with a boundedness requirement as $\xi \to - \infty$ to yield a particular solution of form $p(\xi) = C e^{\xi}$ whose coefficient can be determined by plugging into Eq.(\ref{peqxineg}). Piecing together the solutions on each half-line using a continuity condition, we find
\begin{align*}
    p(\xi) &= \frac{c \gamma}{2 (c+1) (c \gamma \tau_q + 1)} \left[ e^{\xi} + \left( e^{-\xi/(c \tau_q)} - e^{\xi} \right) H(\xi) \right].
\end{align*}
Finally, we recall $(I_u(\xi,t), I_q(\xi,t)) = (\delta(t-t_0),0)$ in Eq.~(\ref{eqn:asymptotic_ode}) for this particular scenario and integrate with respect to time to find
\begin{align}
\nu (t) = \frac{2 (c+1)(c \gamma \tau_q + 1)^2}{2 \theta (c \gamma \tau_q + 1)^2 - (1- \gamma) \gamma \tau_q} H(t - t_0). \label{nuglob}
\end{align}
We demonstrate this prediction and compare to numerical simulations in Fig.~\ref{fig3}B. Note, we generally underestimate the degree to which the front is advanced because this is a global stimulus and so higher order nonlinear effects further speed along the front while it remains away from equilibrium. However for sufficiently small $\ep$, the linear approximation is reasonably close. Taking the limit of no synaptic depression $\gamma \to 1^-$, we see we recover
\begin{align*}
    \lim_{\gamma \to 1^-} \nu(t) = \frac{c+1}{\theta} H(t- t_0) = \frac{H(t - t_0)}{2 \theta^2},
\end{align*}
which was derived previously in \cite{KILPATRICK2012}. Note well that because we are not considering persistent stimuli here, the dependence upon $\nu$ on the right-hand-side of Eq.~(\ref{eqn:asymptotic_ode}) does not come into effect and so it is straightforward to solve the equation by integration.  
As noted, spatially global stimuli perturb waves in ways that are not always well captured by linear theory, but more spatially localized stimuli have a more modest effect on wave position, which can also be captured well by our linear theory as we now show.

\subsection{Pulse Wave Response}
Traveling front solutions are unaffected by activity ($I_u$) perturbations behind their leading edge ($\xi < 0$) as demonstrated above. On the other hand, traveling pulses have both a front and back which can be perturbed by weak inputs. Moreover, spatially localized inputs that may model the position of a visual object interact with wave position in ways that depend on their width as well as strength. To better understand the dynamics of such interactions as a means of building a theory of visual object tracking, we begin by deriving the nullspace of the adjoint operator in the case of a traveling pulse. Taking a Heaviside nonlinearity in Eq.~(\ref{eqn:nullspace}) we find
\begin{subequations} \label{eq:pulnull}
\begin{align}
	c \frac{dv}{d \xi} + v =& \delta (\xi) \frac{Q(0)}{|U'(0)|} \cdot \left[ \int_{\RR} w(y)v(y) dy - \beta p(0) \right] \nonumber \\
		& + \delta(\xi + \Delta) \frac{Q(-\Delta)}{|U'(-\Delta)|} \cdot \left[ \int_{\RR} w(y + \Delta)v(y) dy - \beta p(-\Delta) \right], \label{eq:pullnulla} \\
	c\tau_q \frac{d p}{d \xi} + p =& \left[ H(\xi + \Delta) - H(\xi) \right] \cdot \left[ \int_{\RR} w(\xi - y) v(y) dy - \beta p(\xi) \right], \label{eq:pullnullb}
\end{align}
\end{subequations}
where again the right-hand-side of Eq.~(\ref{eq:pullnulla}) is singularized due to the derivative of the step nonlinearities $f(U(\xi) - \theta) = H(\xi + \Delta) - H(\xi)$ such that
\begin{align*}
    \delta (\xi + \Delta) - \delta (\xi) = \frac{d}{d \xi} \left[ H(\xi + \Delta) - H(\xi) \right] = \frac{d}{d \xi} H(U(\xi) - \theta) = H'(U(\xi) - \theta) U'(\xi).
\end{align*}
As such, it has the form $c v'(\xi) + v(\xi) = A \delta(\xi) + B \delta(\xi + \Delta)$ suggesting the following ansatz $v(\xi) = \frac{1}{c} \cdot \left[ A_0 H(\xi)e^{-\xi/c} + A_{-\Delta} H(\xi+\Delta)e^{-(\xi + \Delta)/c} \right]$ whose coefficients satisfy the linear system
\begin{subequations}
\begin{align}
c A_0 &= \frac{Q(0)}{|U'(0)|} \cdot \left[ A_0 G(0) + A_{-\Delta} G(- \Delta )  - \beta c p(0) \right], \\
   c A_{-\Delta} &= \frac{Q(-\Delta)}{|U'(-\Delta)|} \cdot \left[ A_0 G(\Delta ) + A_{-\Delta} G(0) - \beta c p(-\Delta) \right],
\end{align}
\label{eqn:pulse_consistency}
\end{subequations}
where we have defined the integral function
\begin{align*}
    G(z) = \int_{z}^{\infty} w(y) e^{-(y - z)/c} dy.
\end{align*}
To solve this linear system, we need to determine $p(\xi)$ to constrain its values at $\xi = 0, - \Delta$. Of course, recall $(v,p)^T$ is a vector eigensolution and so can be freely scaled. As such, we should expect the two equations in Eq.~(\ref{eqn:pulse_consistency}) are linearly dependent, so we need only solve one.

\begin{figure}[t!]
\centering
\includegraphics[width=1\textwidth]{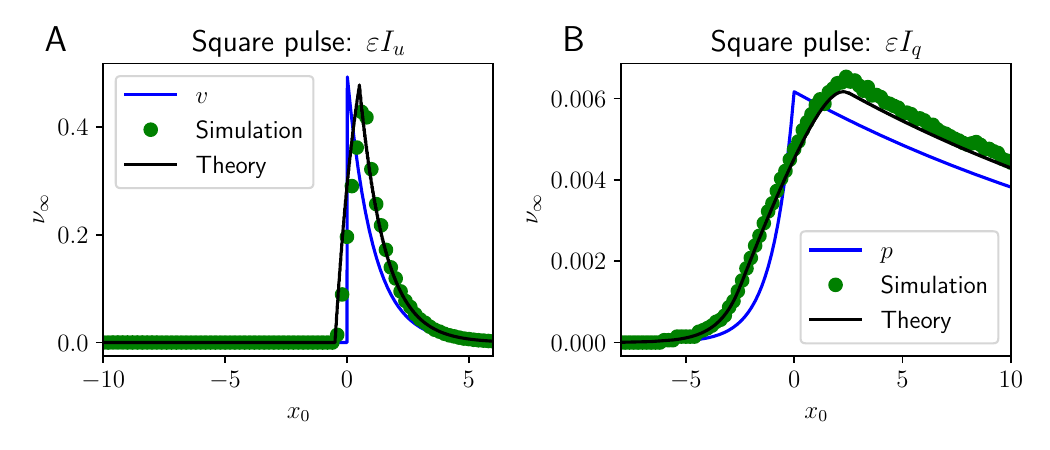}
\caption{
    {\bf Response of traveling pulse to localized pulsatile square stimulus.}
    Stimuli $\ep I_j(x,t) = \ep \bar{I}_j \delta(t) P(x,x_0,\Delta_x)$ ($j = u,q$) are centered in space $P(x, x_0, \Delta_x) = H(x- x_0 + \Delta x/2) - H(x-x_0 - \Delta_x/2)$ at $x_0$ with width $\Delta x = 1$ and presented at time $t=0$ either to ({\bf A}) the neural activity variable: $\bar{I}_u = 1$  and $I_q \equiv 0$; or ({\bf B}) the synaptic efficacy variable: $\bar{I}_q = 1$ and $\bar{I}_u =0$. Asymptotic theory (black line) approximates the results from numerical simulation (green dots) very well. Note that the spatially square pulse input shifts and widens the core null space function (blue curve), which describes how the wave filters forcing.
    Other parameters are $\theta = 0.2$, $\gamma = 1/6$, $\tau_q= 20$, and $\ep = 0.05$.
}
\label{fig4}
\end{figure}

We thus next examine Eq.~(\ref{eq:pullnullb}) in pieces. First, along $\xi <  - \Delta$, we find a simple homogeneous equation $c \tau_q p'(\xi) = -p(\xi)$ whose solution must be bounded as $\xi \to - \infty$, so $p(\xi) = 0$ on this portion of the domain implying $p(-\Delta) = 0$.
Then on $-\Delta < \xi < 0$, we can rewrite the equation as
\begin{align}
    \frac{d p}{d \xi} + \frac{1}{c \gamma \tau_q} p(\xi) = \frac{1}{c^2 \tau_q} \left[ A_0 G(- \xi ) + A_{- \Delta} G(- \Delta - \xi ) \right],
\end{align}
which we can integrate with the boundary condition $p(- \Delta) = 0$ to find
\begin{align}
p(\xi) = \frac{e^{- \xi/(c \gamma \tau_q)}}{c^2 \tau_q} \cdot \left[ A_0 J(\xi, 0) + A_{- \Delta} J(\xi, - \Delta) \right], \label{pxi}
\end{align}
where we have defined
\begin{align*}
    J(\xi, z) = \int_{- \Delta}^{\xi} G(z-y) e^{y/(c \gamma \tau_q)} dy.
\end{align*}
Evaluating Eq.~(\ref{pxi}) at $\xi = 0$ to find the remaining term in the self-consistency Eq.~(\ref{eqn:pulse_consistency}),
\begin{align*}
    p(0) = \frac{A_0}{c^2 \tau_q} J(0, 0) + \frac{A_{-\Delta}}{c^2 \tau_q} J(0, - \Delta)
\end{align*}
Substituting into the consistency conditions Eq.~(\ref{eqn:pulse_consistency}), we have that $A_0$ and $A_{-\Delta}$ satisfy the singular linear system
\begin{align*}
    \left[ \frac{|U'(0)|}{Q(0)}c + \frac{\beta}{c \tau_q} J(0,0) - G(0) \right] \cdot A_0 + \left[ \frac{\beta}{c \tau_q}J(0, - \Delta) - G(- \Delta) \right] \cdot A_{- \Delta} =& 0 \\
    - G(\Delta) A_0 + \left[ \frac{|U'(- \Delta)|}{Q(- \Delta)} c - G(0) \right] \cdot A_{- \Delta} = & 0.
\end{align*}
Ignoring the trivial solution $A_0 = A_{- \Delta} = 0$ and fixing the first entry $A_0 = 1$, we may solve this singular equation by satisfying the second equation to find
\begin{align}
    A_{- \Delta} = \frac{G(\Delta) \cdot Q(-\Delta)}{c | U'(- \Delta)| - G(0) \cdot Q(- \Delta)}, \label{Amdel}
\end{align}
so by defining $A_{- \Delta}$ as in Eq.~(\ref{Amdel}) and $A_0 = 1$, we find the one-dimensional nullspace of ${\mc L}^*$ is spanned by $(v(\xi), p(\xi))^T$ where
\begin{subequations} \label{vpsol}
\begin{align}
    v(\xi) &= H(\xi) e^{- \xi / c} + A_{- \Delta} H( \xi + \Delta) e^{- (\xi + \Delta)/c}, \\
    p(\xi) &= \frac{e^{- \xi/ (c \gamma \tau_q)}}{c^2 \tau_q} \cdot \left[ J(\xi , 0) + A_{- \Delta} J (\xi , - \Delta) \right].
\end{align}
\end{subequations}
To study the impact of spatially locaized and temporally pulsatile stimuli on traveling pulses, we again use Eq.~(\ref{eqn:asymptotic_ode}) to formulate an approximation to the phase advance of the pulse from an abrupt input at a single point in time,
\begin{align}
    \nu(t) = - \ep \frac{\int_{x_0 - \Delta x/2}^{x_0 + \Delta x/2} \left[ \bar{I}_u \cdot v(\xi) + \bar{I}_q \cdot p(\xi) \right] d \xi}{\int_{\RR} U'(\xi) v(\xi) + \tau_q Q'(\xi) p(\xi) d \xi} H(t - t_0). \label{pulnu}
\end{align}
Integrating against both perturbations of the neural activity of amplitude $\bar{I}_u$ and synaptic efficacy of amplitude $\bar{I}_q$ variables, we adjust the position $x_0$ of the stimulus relative to the pulse and see the approximation agrees well with numerical simulations (Fig.~\ref{fig4}). Note that the effect of inputs is stronger (See axis scaling on Fig.~\ref{fig4}A vs \ref{fig4}B) when applied to the neural activity variable ($\bar{I}_u > 0$ and $\bar{I}_q \equiv 0$).
Taking the width of the stimulus to zero $\Delta x \to 0^+$ and its amplitude arbitrarily strong ($\bar{I}_u \to \infty$ or $\bar{I}_q \to \infty$), the input approaches a delta distribution in space and the integral in Eq.~(\ref{pulnu}) is proportional to the nullspace as a function of $x_0$. Generally, we find contributions to the phase advance of the wave are quite weak when originating at the back (threshold crossing) of the pulse
($A_{-\Delta} \approx 0$) as can be seen by evaluating Eq.~(\ref{Amdel}) across a wide range of parameters. Thus, pulses are only substantially perturbed by inputs at their front.

Our experiments for simple stimuli (spatially homogeneous inputs to fronts, Fig.~\ref{fig3}; and square wave pulses to traveling pulses, Fig.~\ref{fig4}) result in trivial differential equations we can solve by integrating due to the stimulus being localized to a single point in time.
When stimuli varying non-trivially in space and time, we must heed the nonlinear nature of the asymptotic approximation of the wave phase given by the differential Eq.~(\ref{eqn:asymptotic_ode}). 
In the next section, we demonstrate that the recurrence represented in this nonlinear equation allows us to better approximate the effects of persistent as well as periodic stimuli providing a simplified theory for predicting the entrainment of traveling waves to localized moving stimuli.

\section{Entrainment to Moving and Apparently Moving Stimuli} \label{sec:entrainment}

We now describe an asymptotic theory of the encoding of moving objects by the position of perturbed traveling waves. In the absence of stimulation, Eq.~(\ref{eqn:model}) supports traveling wave solutions that move at a fixed {\em natural} wavespeed. 
A wave's leading edge is interpreted as the visually encoded position of a moving object. 
When objects move at speeds other than the natural wavespeed, external stimulation must be able to shift the phase and/or increase the speed of the wave to appropriately encode the position. Sensory input is modeled as a spatially localized but moving stimulus $I_u$ with a different speed. We will say such a stimulus {\em entrains} a wave if the wave's speed is altered so as to match that of the stimulus~\citep{ALAMIA2023}.

Our wave response approximation Eq.~(\ref{eqn:asymptotic_ode}) can be used to predict conditions for the entrainment of traveling pulses to both persistent and moving stimuli as well as intermittent flashing stimuli meant to mimic the forcing common to the \textit{apparent motion illusion}.
As opposed to the theory developed previously for the tracking of persistent and moving stimuli~\citep{FOLIAS2005,WU2008B}, flashing stimuli that hop require a periodic map-based theory to determine conditions on speed and input amplitude for perceived motion. We begin by examining a perturbative theory of tracking persistent and moving stimuli first.

\subsection{Front Entrainment}
For simplicity, and to demonstrate the techniques we will use to determine entrainment, we first consider a moving step stimulus defined
\begin{equation}
\varepsilon I_u(x,t) = \varepsilon H\big[(c+\Delta_c)t - x \big].
\label{eqn:moving_heaviside}
\end{equation}
applied to a traveling front solution to Eq.~(\ref{eqn:model}).
For the purposes of this calculation, we consider the propagation of the step at $x = (c+ \Delta_c)t$ as the {\em location} of this stimulus. Note, we will take the difference of the input speed $c + \Delta_c$ and natural wavespeed $c$ to be positive $\Delta_c>0$ so the wave must catch up to the input. We are concerned with deriving conditions on the strength $\ep$ and speed offset $\Delta_c$ of the input that allow for the perturbed wave's speed to match that of the input (entrainment).
When the input leads the wave, it will speed the wave up beyond its natural speed, and if the stimulus is sufficiently strong and slow, the perturbed wavespeed will catch up to that of the input, so the front lags the stimulus at a distance that remains bounded in time. Stimuli that are too fast and/or weak, however, will outrun the wave, so the distance between them grows in time (and entrainment fails).
Analyzing Eq.~(\ref{eqn:asymptotic_ode}), we will determine the relation between input parameters and boundaries of entrainment as well as the distance entrained fronts lag inputs.

For inputs that drive only the neural activity variable of traveling fronts, we find the associated component of the nullspace filtering the input is $v(\xi) = H(\xi)e^{-\xi/c}$. Note, the wave will only be affected by inputs ahead of the front. Moreover, upon defining the scaling factor
\begin{align*}
    K \equiv \langle U', v \rangle + \tau_q \langle Q', p \rangle
\end{align*}
that appears in the denominator of Eq.~(\ref{eqn:asymptotic_ode}), we can simplify the expression for $\nu'(t)$. Now, upon assuming the input is initially ahead of the front, and recalling $\xi = x - ct$, so $H(ct- x + \Delta_c t) = H(- \xi + \Delta_c t)$, we have
\begin{align*}
    K \frac{d \nu}{dt} &= \int_{\RR} v(\xi) I_u(\xi + \ep \nu) d \xi = \int_0^{\infty} H(-(\xi + \ep \nu - \Delta_c t)) e^{- \xi/c} d \xi = c \left( 1 - e^{(\ep \nu - \Delta_c t)/c} \right)
\end{align*}
Applying the change of variables $y(t) = \varepsilon \nu(t) - \Delta_c t$, we can define $y(t) < 0$ as the time-dependent lag of the front behind the stimulus, and we find 
\begin{align}
    K \frac{y' + \Delta_c}{\varepsilon} &= c \cdot\left( 1  -  e^{y/c} \right), \nonumber \\
        \frac{dy}{dt} &=  - \Delta_c + \frac{\ep c}{K} \cdot \left( 1 - e^{y/c} \right)\label{yfde}
\end{align}
Since this is now an autonomous differential equation and assuming the stimulus is ahead of the front ($y<0$), we can derive a corresponding fixed point $y_{\infty} < 0$ under the condition $\Delta_c < \frac{\ep c}{K}$. The equilibrium is specified by the equation
\begin{align}
    \Delta_c = \frac{\ep c}{K} \left( 1 - e^{y_{\infty}/c} \right) \ \ \Rightarrow \ \ y_\infty = c \log \left(1 - \frac{\Delta_c K}{\varepsilon c} \right). \label{yinffront}
\end{align}
When $\Delta_c \geq \frac{\ep c}{K}$, the solution defined in Eq.~(\ref{yinffront}) does not exist, and the stimulus speed ($c + \Delta_c$) is too fast for the wave to entrain. Eq.~(\ref{yfde}) can also be solved directly, yielding
\begin{align*}
    y(t) = c \log \frac{\ep c - \Delta_c K}{\ep c + \left( (\ep c - \Delta_c K) e^{- y_0/c} - \ep c\right) \cdot \exp \left[ - \frac{\ep c - \Delta_c K}{c K} \cdot t\right]},
\end{align*}
where $y(0) = y_0$. In the limit as $t \to \infty$, we find $y(t) \to y_{\infty}$ as expected.

\subsection{Localized Stimulation of Fronts and Pulses}

Our theory describes visual object inputs as localized in space, which can for instance be associated with square waves with some finite width $\Delta_x$ and speed displacement $\Delta_c$ beyond the natural wavespeed. 
Specifically, we define our stimulus by
\begin{align}
    \varepsilon I_u(\xi, t) &=
        \varepsilon \left[ H\left( -(\xi - \Delta_c t) \right) - H\left(- (\xi +\Delta_x - \Delta_c t)\right)\right]. \label{sqwave}
\end{align}
Taking $\Delta_x \to \infty$, we can recover the traveling front stimulus as in the previous section.

Consider the effect the input has upon the front at different displacements. 
When the front of the stimulus is ahead of the front's leading edge, but the active region of Eq.~(\ref{sqwave}) still contains the front ($ \Delta_c t  \in [0, \Delta_x]$), only the part of the stimulus ahead of the front will be filtered by the adjoint nullspace, the same as before. 
Thus, only the portion of the stimulus ahead of or containing the leading edge of the front has any effect on the location of the leading edge of the pulse. Either a fixed point is reached, or the lag becomes larger than the width of the square stimulus and the impact of the stimulus on the front weakens further. We expect the front will then lose the opportunity to entrain to the stimulus at this point, implying entrainment generally must occur before the entire stimulus slips ahead of the leading edge. When solving for the corresponding fixed point of Eq.~(\ref{yfde}), this generates the stricter entrainment condition
\begin{equation}
    \Delta_c < \varepsilon \frac{c}{K}(1 - e^{- \Delta_x/c}).\label{eqn:pulse_threshold}
\end{equation}
Stability of the fixed point described by Eq.~(\ref{yinffront}) can be determined in the same way as before showing it is stable.

\begin{figure}[t!]
\begin{center} \includegraphics[width=\textwidth]{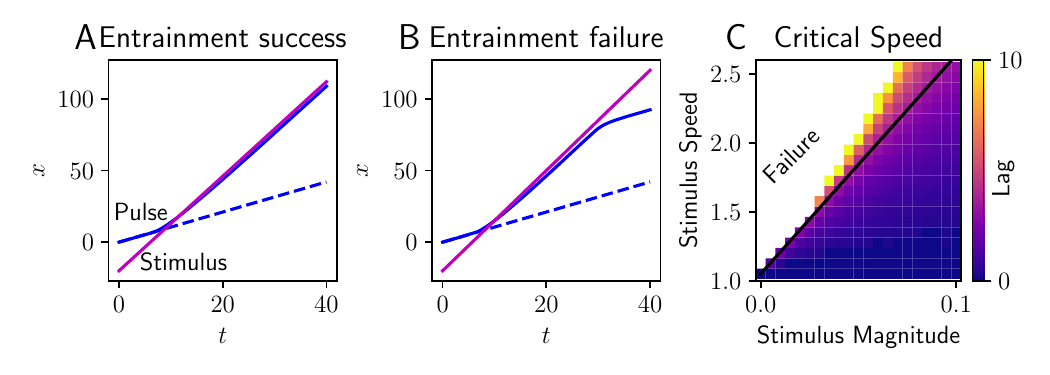} \end{center}
\caption{{\bf Entrainment of traveling pulses to propagating square pulses.} ({\bf A}) Spatiotemporal location of the leading edge of a traveling pulse (solid blue) perturbed by a moving square input (magenta) is ahead of that of the unperturbed pulse (dashed blue). Stimulus has magnitude $\ep = 0.1$ and speed $c + \Delta_c = 3.3$ (compare to natural wavespeed $c \approx 1.051$). Pulse remains entrained indefinitely. ({\bf B}) When stimulus speed $c + \Delta_c = 3.5$ is too large, the stimulus eventually slips off the pulse, which relaxes to its original speed. 
{\bf C.} Numerical simulations reveal a (white) region of failed entrainment when the stimulus speed is too large for a given magnitude $\ep$. Otherwise the colored region indicates the lag $y_{\infty}$ of entrained pulses whose stimulus speeds are not too large. Our first order approximation of the entrainment boundary (Eq.~(\ref{eqn:pulse_threshold}), black line) matches well. Neural field model parameters are $\theta = 0.2$, $\gamma = 1/6$, $\tau_q = 20$ and stimulus width $\Delta_x = 10$.
}
\label{fig5:entrainment}
\end{figure}

Boundaries on entrainment for traveling pulses are determined similarly, except that the activity nullspace term has the form $v(\xi) = H(\xi)e^{-\xi/c} + A_{-\Delta} H(\xi + \Delta)e^{-(\xi + \Delta)/c}.$ 
However, the slow timescale $\tau_q \gg 1$ of the synaptic efficacy results in $|A_{-\Delta}| \ll 1$ across a wide range of parameters. 
This again gives us an entrainment threshold described by Eq.~(\ref{eqn:pulse_threshold}) and a corresponding lag described by Eq.~(\ref{yinffront}), where the pulse speed $c$ and factor $K$ are now determined by the roots of Eq.~(\ref{pulsewidspeed}) and the appropriate inner product, respectively. As expected, in simulations, a stimulus with speed close enough to the natural speed of the pulse will entrain it (Fig.~\ref{fig5:entrainment}A), but if the stimulus speed is too far from the natural speed of the pulse, the pulse slips behind the stimulus indefinitely (Fig.~\ref{fig5:entrainment}B). For weak stimuli, we find that the theory developed here accurately predicts the boundary between entrainment failure and success accurately for corresponding simulations (Fig.~\ref{fig5:entrainment}C).

\subsection{Pulse Entrainment to Apparent Motion} \label{sec:pulse_entrainment}
Apparent motion is modeled here by flashing stimuli whose positions move between but not during flashes (i.e., on periods of the stimulus). In this case, the driven solution does not converge to a proper traveling wave, but a solution that is periodic under a fixed spatial shift given by the movement of the flashing stimuli. Dividing the time domain into segments of length $T$, the period of the forcing stimulus, we define the length of the on phase of length $T_\text{on}< T$, during which there is a square pulse input, and off phase of length $T_\text{off} = T - T_\text{on}$, during which there is no input.
The stimulus is advanced by $(c+\Delta_c)T$ between on phases, and we thus call $\Delta_c$ the speed offset of the input. 
Such a stimulus can be expressed using indicator functions by
\[
    \varepsilon I_u(x, t) = \varepsilon {\mc I}_{(x^* - \Delta_x, x^*)} \left(x - \left\lfloor \frac{t}{T} \right \rfloor (c + \Delta_c) T \right)
    \cdot
    {\mc I}_{(0, T_\text{on})}\left( t - \left\lfloor \frac{t}{T} \right\rfloor T \right),
\]
where we have defined the indicator function ${\mc I}_A(x)$ of a subdomain $A$ with argument $x$ as
\[
{\mc I}_A(x) = \begin{cases}
    1, & \text{ if } x \in A, \\
    0, & \text{ else}.
\end{cases}
\]
Changing variables $(x,t)$ to $(\xi,t)$ and using our simplifying approximation $A_{-\Delta} \approx 0$ and substituting into Eq.~(\ref{eqn:asymptotic_ode}) and making the change of variables $y = \ep \nu + ct$, we can approximate the movement of the leading edge $y$ of the front when it is inside the stimulus,
\begin{align*}
\frac{dy}{dt} = c + \frac{\varepsilon c}{K} \left( 1 - e^{-\Delta_x/c} e^{y/c} \right),
\end{align*}
which can be solved assuming $y(0) = y_0$ is the location of the wave front at time $t=0$, to find
\begin{align*}
    y(t) = - c \log \left[ \frac{\ep}{K + \ep} e^{- \Delta_x/c} + \left( e^{- y_0/c} - \frac{\ep}{K + \ep} e^{- \Delta_x/c} \right) e^{- (K + \ep) t/K} \right] \equiv F(t; y_0),
\end{align*}
so the function $F(t; y_0)$ describes how the input maps $y$ forward from $y_0$ after a time $t$. During the off phase, $y$ will simply advance according to the natural wavespeed, so the approximate solution in the first period $T$ is
\begin{equation}
    y(t) = \begin{cases}
			F(t,y_0) , & 0 \le t \le T_\text{on} \\
		F(T_\text{on},y_0) + (t - T_\text{on})c, & T_\text{on} \le t \le T.
	\end{cases} \label{eqn:apparent_motion_solution}
\end{equation}
Subtracting off $(c+\Delta_c)T$, we can determine where the wave front $y$ will be at time $t = T$ relative to the newly flashed stimulus
\begin{align*}
    y_1 = F(T_{\rm on}, y_0) - c T_{\rm on} - \Delta_c T.
\end{align*}
Thus, we can inductively define a map describing how the next relative location of the wave front $y_{n+1}$ depends on the current location $y_n$ as
\begin{align}
    y_{n+1} = F(T_{\rm on}, y_n)  - c T_{\rm on} - \Delta_c T. \label{flashmap}
\end{align}

\begin{figure}[t!]
\centering
\includegraphics[width=1\textwidth]{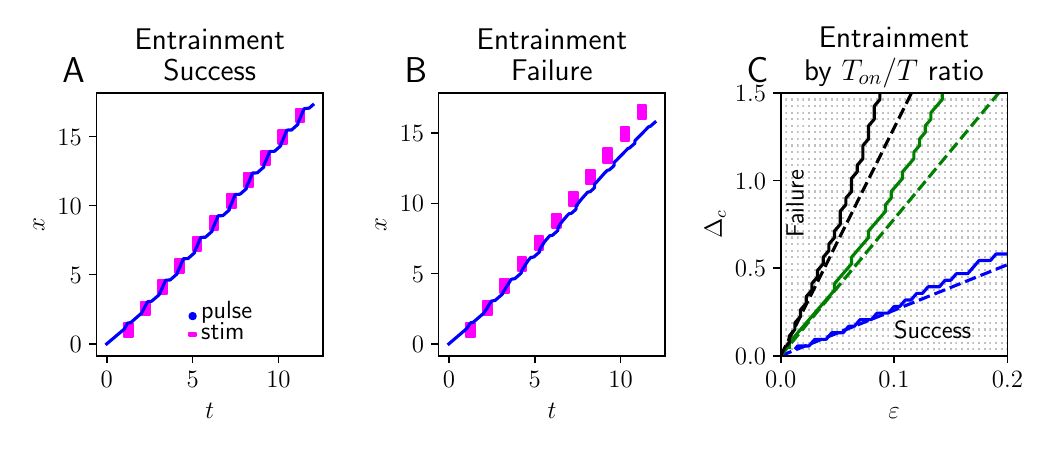}
\caption{
    {\bf Wave entrainment to apparently moving stimuli.} ({\bf A}) The leading edge (blue line) of a traveling pulse becomes entrained to an intermittent, moving, and flashing stimulus (magenta indicates regions where stimulus is non-zero). The difference between the baseline wavespeed ($c \approx 1.051$) and the effective speed of the forcing stimulus is
     $\Delta_c = 0.5$; its magnitude is $\ep = 0.2$; its width and starting position are $\Delta_x = 1$ and $x^* = 0$; on and off phases are $T_{\rm on} = T_{\rm off} = 0.5$ for a total stimulus period of $T = 1$. ({\bf B}) Weakening the stimulus magnitude to $\varepsilon = 0.12$, the traveling pulse fails to entrain, and slips ever further behind the stimulus.
    {\bf C.} Asymptotic approximation to the entrainment boundary (dashed line) $\Delta_c = \ep \frac{T_{\rm on}}{T} \frac{c}{K}$ is well matched to results of numerical simulations (solid) separating the domain of entrainment (lower right) from entrainment failure (upper left). Increasing the ratio $T_{\rm on}/T_{\rm off}$ enlarges the domain of entrainment. Blue: $T_{\rm on} / T_{\rm off} = 0.1/0.5 = 0.2$; Green: $T_{\rm on} / T_{\rm off} = 0.5/0.5 = 1$; Black: $T_{\rm on} / T_{\rm off} = 0.5/0.1 = 5$. Stimulus width is $\Delta_x = 10$; all other model parameters are same.
    Throughout, neural field model parameters are $\theta = 0.2$, $\gamma = 1/6$, and $\tau_q = 20.0$.
}
\label{fig:apparent_motion_entrainment}
\end{figure}

We shall consider wave front entrainment to the periodic, moving, and flashing stimulus possible when the map Eq.~(\ref{flashmap}) has a stable fixed point
described by the condition
\begin{align*}
    y^* = F(T_{\rm on}, y^*) - c T_{\rm on} - \Delta_c T,
\end{align*}
which can be solved to find
\begin{align}
    y^* &= \Delta_x - c \log\left[ \frac{\ep}{K+ \ep} \cdot \frac{e^{T_{\rm on}} - e^{-\ep T_\text{on}/K}}{e^{-\Delta_c T/c} - e^{-\ep T_\text{on}/K}} \right]. \label{perentfp}
\end{align}
The denominator of the argument in the logarithm above is positive, and the solution is defined if $\Delta_c < \ep \frac{T_{\rm on}}{T} \frac{c}{K}$. Linearizing the map Eq.~(\ref{flashmap}) and plugging in the solution Eq.~(\ref{perentfp}), we find its stability is determined by the single eigenvalue $\lambda = \exp \left[ \frac{\Delta_c T}{c} - \frac{\ep T_{\rm on}}{K} \right]$, so $\lambda \in (0, 1)$ if $\Delta_c < \ep \frac{T_{\rm on}}{T} \frac{c}{K}$, which implies that the fixed point defining periodic entrainment is stable whenever it exists.

Our theory is consistent with the result of numerical simulations of the response of traveling waves to periodic, moving, and flashing stimuli. Flashing stimuli that are sufficiently strong and that travel at a speed that is not too much faster than the natural wavespeed will entrain the traveling wave (Fig.~\ref{fig:apparent_motion_entrainment}A). Tracking the leading edges of these flash-forced traveling pulses reveals that the stimulus speeds up the wave while it is on, and the pulse speed relaxes to its natural value between flashes. Over time, the average speed of the forced solution matches that of the average speed $c + \Delta_c$ of the forcing stimulus. On the other hand, if (a) the distance traveled with each hop between flashes is too large or (b) the stimulus is too weak or short, then the pulse will not be sped up enough during the on phase of the period. As a result, the pulse will eventually lag further and further behind the forcing stimulus as time goes on (Fig.~\ref{fig:apparent_motion_entrainment}B). The combined necessity of having a forcing stimulus whose speed difference is not too large ($\Delta_c$), whose magnitude is not too weak ($\ep$), or whose on phase is not too short ($T_{\rm on}$) is all contained in the entrainment boundary inequality $\Delta_ c < \ep \frac{T_{\rm on}}{T} \frac{c}{K}$. Indeed, we see that this boundary well approximates the boundary we can determine from numerical simulations (Fig.~\ref{fig:apparent_motion_entrainment}C).

Thus we find that the entrainment of neural activity waves in a model of sensory cortex can be described by a relatively accessible theory. Akin to results from past work on phase response theory, we find that stimuli whose speed is close enough to the natural speed can indeed entrain waves. This applied not only to persistent traveling stimuli but also flashing stimuli, suggesting a neuromechanistic theory for the generation of apparent motion illusions in visual cortex.

\section{Discussion}
\label{discussion}

In this paper, we have developed and analyzed a neuronal network model describing the stimulus-response relationships of traveling activity waves to moving inputs. Our model incorporates negative feedback in the form of short-term synaptic depression, which attenuates activity at the back of the waves, generating traveling pulses when it is strong enough. The nonlinear model of synaptic depression can be derived directly from spiking models~\citep{TSODYKS1998}, and is more physiologically motivated than heuristic linear negative feedback models~\citep{PINTO2001}.
In the absence of inputs, we can explicitly derive conditions for the emergence of traveling fronts and pulses, along with their speed and width. Building on past work~\citep{KILPATRICK2010B}, we have identified conditions for the coexistence of stable progressing and receding fronts. Using asymptotic theory, we have derived a general formula describing how traveling fronts and pulses respond to generalized external input. Our formula can consider inputs either to the neural activity or synaptic efficacy variables. This framework was then used to perturbatively quantify the effect of different moving stimuli on the phase of traveling waves. The nonlinear differential equation accounts for how past perturbations have shifted the wave when considering the effect of future inputs.

Weak external stimuli shift the position of traveling waves relative to their natural position, which changes linearly in time as they propagate. Extending prior work~\citep{KILPATRICK2012}, we have demonstrated how the impact of inputs both on the leading and trailing edge of a traveling pulse impacts the wave phase, though in general the effect of the trailing edge is weak compared to the leading edge due to the exponential dependence on distance. The key predictions we have made using this theory focus on entrainment phenomena, whether a traveling wave will lock to an external moving stimulus traveling at a different speed. The natural speed of the wave in addition to the strength and relative speed of the input, rather than the width of the stimulus, are most important for making these predictions when inputs are persistent. Indeed propagating neural activity patterns in medial temporal (MT) cortex have been shown to encode object motion direction and speed, suggesting these waves play an important role in neural information processing~\citep{TOWNSEND2017}. However, given that neural activity waves can propagate spontaneously or can persist after an initial stimulus has been removed~\citep{XU2007}, such persistent activity could represent illusory sensory stimuli in the absence of inputs. Our theory demonstrates that propagating waves can be perturbed to match their speed to intermittent stimuli that jump locations between flashes, providing a simple quantitative measure to determine whether such entrainment is expected based on the strength and temporal properties of the input.

Our reduced equations describe the movement of a traveling wave's leading edge relative to the location of a stimulus, allowing us to identify both the fixed points and the transient dynamics of the edge in the coordinate frame of the moving input.
For persistent and moving inputs, this corresponds to a solution to a nonlinear differential equation, which we can obtain explicitly.
For intermittent flashing stimuli, we frame the entrainment problem as a fixed point analysis of a map, describing the updated location of the edge after each temporal period of the stimulus.
Intermittent stimuli must be stronger than corresponding persistent stimuli to entrain wave motion, consistent with observations that real motion is encoded more readily than apparent motion~\citep{MERCHANT2003}.
Our asymptotic techniques allow us to derive a simple formula for the entrainment boundary, which intuitively quantifies the increasing relative speeds that can be entrained as the on phase of the flash grows longer.

The response of traveling waves to external stimuli has been measured in vivo and in vitro~\citep{WU2008}, demonstrating inputs speed up~\citep{RICHARDSON2005}, displace or annihilate~\citep{GAO2012}, or even switch the direction~\citep{PANG2020} of waves. These response properties reflect a highly dynamic and spatiotemporal form of input processing in sensory cortices~\citep{ERMENTROUT2001,MULLER2018}. Disentangling the role of synaptic network architecture in spatiotemporal processing thus requires the analysis of dynamic and mechanistic neural network models that capture cortical complexity while still being amenable to reductions that identify key system input-output relationships. Our analysis has developed a powerful and reduced understanding of how a spatially structured network can encode apparent motion, a phenomenon originally reported in non-invasive behavioral experiments~\citep{ANSTIS1980,RAMACHANDRAN1986} and more recently shown to correspond to elevations of activity in visual cortex in corresponding regions of perceived motion~\citep{MUCKLI2005,BLOM2020}. Activation of visual stimulus-encoding neurons in the absence of a stimulus may suggest the anticipation a temporarily occluded and moving object's reappearance as in the phenomenon of visual preplay~\citep{EKMAN2017,AITKEN2020}. Our results could guide future behavioral and neurophysiological experiments linking psychophysical thresholds for apparent motion to the coordination of visual cortical activity and external inputs.

A number of modeling assumptions could be relaxed, only with cost to explicit tractability of the model, but without sacrificing the core mechanisms of the model. Waves, their linearizations, and eigenfunctions can all be calculated using alternative numerical methods such as shooting~\citep{ERMENTROUT2010} or Fourier methods in approximating partial differential equation models~\citep{COOMBES2007} in order to carry out a similar analysis for smooth firing rate functions, more sophisticated weight functions, or compact domains. Distinct excitatory/inhibitory populations could be considered or additionally short term facilitation to more fully capture the diversity of static and dynamic models of synaptic transmission and understand how these contribute to the filtering properties of traveling waves. Similar analyses could also be applied to planar neural field models, building on past work which has developed perturbative theory based on a spectral analysis of linearized problems to understanding how the closed boundary of a stationary or moving bump responds to external inputs~\citep{COOMBES2012}.

\backmatter

\section*{Code Availability} All software for carrying out model simulations and for generating figures can be found at the following code repository: \url{https://github.com/shawsa/neural-field-synaptic-depression}

\section*{Acknowledgements}

This work was funded by NIH BRAIN 1R01EB029847 and NSF DMS-2207700.


\end{document}